\documentclass[preprint,11pt]{elsarticle}
\usepackage[pdfborder={0 0 0}]{hyperref}
\usepackage{amsthm,amsmath,amsfonts,latexsym,amssymb}
\usepackage{graphics,fancyhdr,graphicx}
\usepackage[ruled,linesnumbered]{algorithm2e}
\usepackage[margin=1in]{geometry}
\usepackage{multirow,booktabs}
\usepackage[title]{appendix}
\usepackage[table]{xcolor}
\usepackage{listings}
\usepackage{tabularx}
\usepackage{placeins}
\usepackage{comment}
\usepackage{lineno}
\usepackage{lscape}
\usepackage{bm}
\usepackage{caption}
\usepackage{subcaption}
\usepackage{float}
\usepackage{soul}
\usepackage{setspace}
\usepackage{multirow}
\hypersetup{
    colorlinks = true,
    urlcolor   = blue,
    citecolor  = custom-blue,
}
\usepackage{textgreek}
\usepackage[utf8]{inputenc}
\usepackage[T1]{fontenc}
\biboptions{numbers,sort&compress}
\definecolor{listinggray}{gray}{0.9}
\definecolor{lbcolor}{rgb}{0.9,0.9,0.9}
\definecolor{Darkgreen}{RGB}{0,100,0}
\usepackage{amssymb}
\usepackage{lineno}
\usepackage{comment}

\def \yb{\bm{y}}
\emergencystretch=\maxdimen
\DeclareMathOperator{\real}{\mathbb{R}}
\begin{document}
\abovedisplayskip=6.0pt
\belowdisplayskip=6.0pt

\begin{frontmatter}

\title{Real-Time Prediction of Gas Flow Dynamics in Diesel Engines using a Deep Neural Operator Framework}
\author[1]{Varun Kumar}
\author[2]{Somdatta Goswami}
\author[3]{Daniel Smith}
\author[1,2]{George Em Karniadakis\corref{mycorrespondingauthor}}

\cortext[mycorrespondingauthor]{Corresponding author. Email: george\_karniadakis@brown.edu}

\address[1]{School of Engineering, Brown University, Providence, RI, USA}
\address[2]{Division of Applied Mathematics, Brown University, Providence, RI, USA}
\address[3]{Cummins Inc., Columbus, IN, USA}

\begin{abstract}
We develop a data-driven deep neural operator framework to approximate multiple output states for a diesel engine and generate real-time predictions with reasonable accuracy. As emission norms become more stringent, the need for fast and accurate models that enable analysis of system behavior have become an essential requirement for system development. The fast transient processes involved in the operation of a combustion engine make it difficult to develop accurate physics-based models for such systems. As an alternative to physics based models, we develop an operator-based regression model (DeepONet) to learn the relevant output states for a mean-value gas flow engine model using the engine operating conditions as input variables. We have adopted a mean-value model as a benchmark for comparison, simulated using Simulink. The developed approach necessitates using the initial conditions of the output states to predict the accurate sequence over the temporal domain. To this end, a sequence-to-sequence approach is embedded into the proposed framework.  The accuracy of the model is evaluated by comparing the prediction output to ground truth generated from Simulink model. The maximum $\mathcal L_2$ relative error observed was approximately $6.5\%$. The sensitivity of the DeepONet model is evaluated under simulated noise conditions and the model shows relatively low sensitivity to noise. The uncertainty in model prediction is further assessed by using a mean ensemble approach. The worst-case error at the $(\mu + 2\sigma)$ boundary was found to be $12\%$. The proposed framework provides the ability to predict output states in real-time and enables data-driven learning of complex input-output operator mapping. As a result, this model can be applied during initial development stages, where accurate models may not be available.
\end{abstract}

\begin{keyword}
diesel engine \sep neural networks \sep non-linear dynamics \sep operator learning \sep uncertainty quantification 
\end{keyword}

\end{frontmatter}

\section{Introduction}
\label{sec:introduction}

Diesel engines are used extensively in heavy-duty applications due to their higher peak torque and thermal efficiency as compared to their gasoline counterparts. However, because these engines emit health-hazardous nitrogen oxides ($NO_{x}$) and particulates, strict emission control limits are placed on them. Trade offs between optimum operating conditions and engine emissions are often made to ensure compliance with regulatory norms. In this context, understanding the operation of combustion engines through model-based engineering has been a popular approach for product development in the automotive industry. Engine manufacturers are continually searching for ways to improve performance by altering the fundamental operational settings to enhance either of these performance indicators. Analytical models have been widely used for simulating the behavior of combustion engines, and several commercial packages exist to enable modeling their behavior \cite{AVLBoost, GTPower, RicardoWave}. The mean-value models for simulating diesel engine gas flow, proposed in \cite{hendricks1986compact, watson1984dynamic}, are based on manifold filling and emptying concept. These models allow simulating the engine behavior by making approximations around engine transient behaviors and have been used extensively in control design and fault diagnosis \cite{kimmich2005fault, wu2011mean, svard2010residual, han1995turbulence}. Although significant progress has been made in improving the prediction capability of engine simulation models, it has been acknowledged that the complexity of diesel engine control based on numerical models rise with the number of independent variables, necessitating the use of multidimensional, flexible, and adaptive add-ons. 

Recent advancements in data- and physics-driven surrogate modeling have shown significant success in their ability to simulate behavior of complex systems \cite{goswami2020adaptive, goswami2021physics}. These models rely on using empirical data for creating representations of real system behavior through analytical means. Deep learning techniques are at the forefront of data-driven modeling paradigms due to their inherent ability to model complex non-linear relationships using labelled datasets \cite{lecun2015deep}.  Additionally, these surrogate models offer fast predictions and this is essential for field deployment. However, the ability of deep learning algorithms to handle real data, which is frequently accompanied by noise, presents additional challenges when modeling internal combustion (IC) engines. 

The objective of the current work is to develop a robust and efficient surrogate model to simulate diesel engine operations using appropriate deep learning techniques. Specifically, we develop a deep operator-based network (referred to as DeepONet herein) to predict the gas flow dynamics of a diesel engine. While the inputs to the network are sensory measurements from field, the ground truth is simulated using a mean-value engine Simulink model\footnote{Software packages from Vehicular Systems by Johan Wahlstr{\"o}m, and Lars Eriksson. \url{http://www.fs.isy.liu.se/Software}}. The developed surrogate model takes into account the independent nature of various parameters that constitute the complete analytical model for IC engines. Additionally, we also carry out a comprehensive study for model uncertainty. Enlisted below are our main contributions through this work:

\begin{itemize}
    \item Demonstrate application of the deep neural operator (DeepOnet) to predict seven output states (intake manifold pressure $P_{im}$, exhaust manifold pressure $P_{em}$, residual gas fraction $x_{r}$, temperature after inlet valve closes at intake completion $T_{1}$, turbo-shaft speed $\omega_{t}$, EGR actuator signal $\tilde u_{egr}$, and VGT actuator signal $\tilde u_{vgt}$),  with a maximum relative $\mathcal L 2$ error of $\approx 6.5\%$ across a $1,000$ seconds  prediction window. We use the existing Simulink model as a ground truth data generator to train our DeepONet model and then predict the output states for unseen input samples.  
    \item \emph{Ability to generate real time predictions in less than a second} with a given set of inputs. Once the DeepONet is trained, generating predictions from the trained model takes minimal time. The ability to generate instantaneous and accurate predictions holds significant advantages for real world implementation on such systems.
    \item Demonstrate an exemplar DeepONet architecture for learning the functional mapping between input and output states for a diesel engine. This mapping is performed based on the engine speed, fueling, EGR valve position, and VGT valve position data as inputs to the DeepONet model without specific knowledge of the governing equations, which at times may not accurately known. 
    \item Identify the uncertainty of the proposed DeepONet model with noisy inputs and determine errors under such conditions. The surrogate model shows an increase in prediction error with increased levels of noise, but this increase is within an acceptable limit. The maximum $\mathcal{L}_{2}$ error calculated with the added noise to input was $\approx 7\%$ for the output state $P_{em}$. 
    \item Determine uncertainty in model estimation through the use of dropout in the network architecture. We quantify the maximum uncertainty that exists in the DeepONet model predictions through the use of an ensemble based approach. The maximum relative $\mathcal{L}_{2}$ error at $2\sigma$ standard deviation from the ensembled mean was found to be 12.6\%, which was approximately 6\% higher than the error reported for $P_{em}$ using the ensemble mean.
\end{itemize}

The remainder of the manuscript is arranged as follows. In section \ref{sec:numerical_simulation}, the Simulink model used for generating the ground truth for training the deep operator network is presented briefly. Minor modifications made to the Simulink model to meet the objectives of the current work are discussed. In section \ref{sec:regression_modeling}, we present a brief overview of the DeepONet architecture and showcase the developed surrogate model for modeling the diesel engine. Section \ref{sec:results} presents the details of the experiments conducted along with the effect of adding noise on prediction accuracy. Section \ref{sec:uncertainty_eval} presents the results for model uncertainty through dropouts used in the branch network. Lastly, in section \ref{sec:summary}, we present a brief summary of our observations and report known limitations for the current methodology.

\section{Numerical simulation of the diesel engine}
\label{sec:numerical_simulation}
\begin{figure}[!t]
  \centering
  \includegraphics[width = \textwidth]{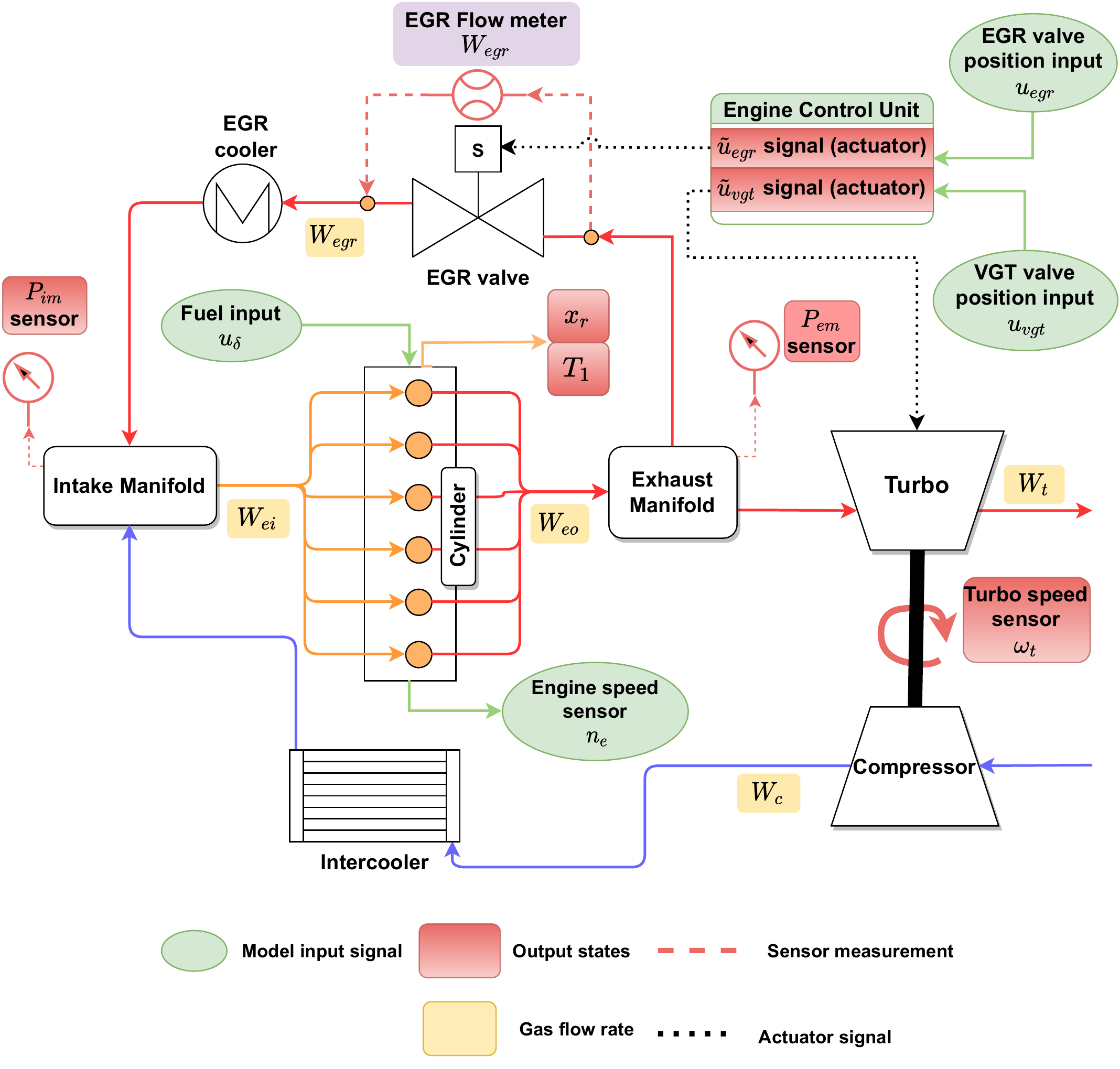}
  \caption{Schematic of various subsystems in the mean-value diesel engine model (modified from \citep{wahlstrom2011modelling}). In this schematic, $n_{e}$, $u_{\delta}$, $u_{egr}$ and $u_{vgt}$ represents the engine speed, the fuel injected into the cylinders per cycle, valve position signals received by the EGR valve and the turbocharger valve actuators, respectively. The terms with $W$ denote mass flow through the component, where $W_{ei}$ represents gas flow rate into the cylinders, $W_{eo}$ represents exhaust gas flow rate into the exhaust manifold, $W_{egr}$ represents EGR gas flow rate, and $W_{c}$\ represents the fresh air flow rate into the intake manifold from the compressor.}
\label{fig:engine_schematic}
\end{figure}

The rapid exchange of air and exhaust gases alongside energy inside a diesel engine presents a challenge in creating representative models for analysis. Mean-value models based on the emptying and filling of manifold volumes have been proposed for simplicity (\cite{hendricks1986compact, watson1984dynamic} ). The engine model by Wahlstr{\"o}m and Ericsson \cite{wahlstrom2011modelling} is one such simplified model and is based on the dynamics of gas flow inside the manifolds, EGR valve, and turbocharger and is the source of simulated data generation in this study. In this section, we briefly discuss the inputs and outputs associated with this model for clarity. Figure \ref{fig:engine_schematic} presents a schematic of the various components associated with Wahlstr{\"o}m's model. The inputs and the outputs of this model are signals recorded using dedicated sensors over a certain period of time. The input signals of this model can be defined by the input vector: $$\text{inputs} : [n_e,\; u_{\delta},\;u_{egr},\;u_{vgt}],$$ where $n_{e}$ represents the engine speed,  $u_{\delta}$ the fuel injected into the cylinders per cycle, $u_{egr}$ and $u_{vgt}$ represent EGR and VGT valve openings that are empirically determined during engine calibration. The output states emanating from this model are defined by the output vector: $$ \text{output\,states}: [P_{im},\; P_{em},\;X_{O_{im}},\; X_{O_{em}},\; \omega_t,\;\tilde u_{egr1},\; \tilde u_{egr2},\; \tilde u_{vgt}],$$ where $P_{im}$ is the input manifold pressure, $P_{em}$ represents the exhaust manifold pressure, $X_{O_{im}}$ represents oxygen mass fraction in intake manifold, $X_{O_{em}}$ the oxygen mass fraction in exhaust manifold, $\tilde u_{egr1,2}$ for EGR actuator dynamics and $\tilde u_{vgt}$ represents VGT valve actuator dynamics. The eight output states are obtained by solving their respective ordinary differential equations (ODEs) by using conventional numerical solvers in Simulink. Output data generated from the Simulink model is used as the ground truth in this work. Interested readers can refer to \cite{wahlstrom2011modelling} for details on the relations of the inputs and the output states, and additional parameters of the Simulink model \cite{Simulink}.

To achieve the goals of our current work discussed in Section \ref{sec:introduction}, we modified the original Simulink model to extract the desired output states. These changes were required to make this work compatible with parameter identification task in future. Additional blocks were added to the Simulink model for extracting the necessary output states. The new output output states extracted from the modified model are represented as $$\text{new output states}: [P_{im},\; P_{em},\;  x_{r},\; T_{1},\; \omega_t, \;\tilde u_{egr},\; \tilde u_{vgt}],$$ where $x_{r}$ represents the residual gas fraction,  $T_{1}$ is the temperature once inlet valve is closed after intake stroke, and $\tilde u_{egr}$ represents the combined EGR valve actuator output obtained through a combination of $\tilde u_{egr1}$ and $\tilde u_{egr2}$ (see equation 39 in \cite{wahlstrom2011modelling}). The input data is collected from a $6$-cylinder heavy-duty truck engine on a test bed. The parameters required for generating the output data from the Simulink model is adopted from \cite{wahlstrom2011modelling}. A sampling frequency of $2$ Hz is used for generating the ground truth solution used for evaluating the accuracy of our surrogate model.

The Wahlstr{\"o}m and Ericsson (WE) engine model employs a set of parameters derived from a least square fit of measured values from the engine laboratory. Hence, in order to adopt this model for different engines, these parameters must be determined. For instance, the Simulink model for EGR valve employs coefficients $c_{egr1}, \, c_{egr2}, \, c_{egr3},$ and $\Pi_{egropt}$, which are determined through empirical curve fitting. Therefore, to simulate and employ the WE model for other applications, additional effort is required to determine the parameters based on the analytical relations. Computing these parameters requires one to solve inverse problems within an optimization loop that increases latency in predictions. Surrogate models serve as cost effective approximations for high-fidelity simulations, allowing for significant computational savings while maintaining solution accuracy. Deep neural operators (DeepONet), introduced in 2019 \cite{lu2021learning}, have been employed effectively as surrogate models for complex physical problems like fracture mechanics \cite{goswami2022physics}, bubble dynamics \cite{lin2021operator}, and electro-convection \cite{cai2021deepm} to name a few. One significant advantage of an operator-based model is its ability to learn nonlinear functional mappings between inputs and outputs based on data. Additionally, the prediction time for a pre-trained DeepONet is a fraction of second, which is a critical requirement for real-time forecasting in field application. The flexibility offered by the DeepOnet allows generalization to different operating conditions such as different ambient temperature and pressure, which is another advantage over the traditional solvers. In the next section, we showcase the proposed operator regression model developed for the mapping of the input conditions to the output states.

\section{Operator regression for Diesel Engine modeling}
\label{sec:regression_modeling}

A deep operator network allows learning a non-linear operator from data and is suited for application where the physics-based models are difficult to ascertain or when generalized results are desired. Diesel engines, with their high frequency dynamic processes are a suitable candidate for using an operator based neural network. The idea of DeepONet is motivated by the universal approximation theorem for operators \cite{chen1995universal}, which states that a neural network with a single hidden layer can approximate accurately any linear/non-linear continuous function or operator. Before we begin learning the solution operators of the parametric diferential equations, we must first distinguish between a function regression and an operator regression. The solution in the function regression approach is parameterized as a neural network between finite dimensional Euclidean spaces: $\mathcal{F}:\mathbb{R}^{d_1}\to\real^{d_1}$, where $d_1$ is the number of discretization points. In operator regression, however, a function is mapped to another function using an operator. With this idea in mind, we put forward the conventional architecture of DeepONet in the first part of this section and later introduce the proposed architecture of the DeepONet specific to this work.

\subsection{The Deep Operator Network (DeepONet)}
\label{subsec:deeponet}

 DeepONet consists of two deep neural networks (DNN): $\mathcal N_1$ (conventionally called the branch net) encodes the input function, $\boldsymbol u$ at $m$ fixed locations (typically called sensors), and $\mathcal N_2$ (termed as the trunk net) inputs the location of evaluation of the solution, $\boldsymbol y$ (trunk net). In a generalized setting, the branch network input can take the shape of the physical domain, initial or boundary conditions, constant or variable coefficients, source terms, and so on, as long as the input function is discretized at $m$ sensor locations. A convolutional neural network (CNN) can be used as the branch net for a regularly spaced discretization of the input function, whereas for a sparse representation, a feed-forward neural network (FNN) or even a recurrent neural network (RNN) for sequential data can be considered. In this work, we have used FNNs to represent both the trunk and the branch networks.

To recognize the theoretical underpinning of a DNN, we consider a neural network with $L$ hidden layers, with the $0$-th layer denoting the input layer and the $(L+1)$-th layer denoting the output layer; the weighted input $\bm z^l_i$ into a $i$\textsuperscript{th} neuron on a layer $l$ is a function of weight $\bm W^l_{ij}$ and bias $\bm b^{l-1}_j$ and is represented as
\begin{equation} \label{eq:weighted_input}
    \bm z^l_i =  \mathcal{R}_{l-1}\left(\sum_{j=1}^{m_{l-1}}\left(\bm W^l_{ij}(\bm z^{l-1}_j) + \bm b^{l}_j\right)\right),
\end{equation}
where $m_{l-1}$ is the number of neurons in layer $l-1$ and $\mathcal{R}_{l-1}\left( \cdot \right)$ represents the activation function of layer $l$. The feed-forward procedure for calculating the output $\bm Y_L$ is expressed as follows based on the aforementioned concepts:
\begin{equation} \label{eq:feedforward}
\begin{split}
       \bm{Y}^L &= \mathcal{R}_L(\bm W^{L+1}\bm{z}^L + \bm b^L)\\
       \bm{z}^L & = \mathcal{R}_{L-1}\left(\bm{W}^{L}\bm{z}^{L-1} + \bm b^L\right)\\
       \bm{z}^{L-1} & = \mathcal{R}_{L-2}\left(\bm{W}^{L-1}\bm{z}^{L-2} + \bm b^{L-1}\right)\\
        &\;\;\;\;\;\;\;\vdots\\
        \bm z^1 &= \mathcal{R}_0\left(\bm W^1\bm{x} + \bm b^1\right),\\
\end{split}
\end{equation}
\noindent where $\bm{x}$ is the input of the neural network. \autoref{eq:feedforward} can be encoded in compressed form as $\bm Y = \mathcal N (x;\bm{\theta})$, where $\bm{\theta} = \left(\bm W, \bm b \right)$ includes both the weights and biases of the neural network $\mathcal N$. Taking into account a DeepONet, $\mathcal N_1$ takes as input the function to denote the input realizations $\bm{U} = \{\bm{u}_1, \bm{u}_2, \ldots, \bm{u}_N\}$ for $N$ samples, discretized at $n_{sen}$ sensor locations such that $\bm{u}_i = \{u_i(\bm x_1), u_i(\bm x_2), \ldots, u_i(\bm x_{n_{sen}})\}$ and $i \in [1,N]$, while $\mathcal N_2$ inputs the location $\bm y = \{\yb_1,\yb_2,\cdots,\yb_p\}=\{(\hat x_1,\hat y_1),(\hat x_2, \hat y_2), \ldots, (\hat x_p,\hat y_p)\}$ to evaluate the solution operator, where $\hat{x}_i$ and $\hat{y}_i$ denote the coordinates $x$ and $y$ of the point $\yb_i$, respectively. Let us consider that the branch neural network consists of $l_{br}$ hidden layers, where the $(l_{br}+1)$\textsuperscript{th} layer is the output layer consisting of $q$ neurons. Considering an input function $\bm u_{i}$ in the branch network, the network returns a feature embedded in $[b_1, b_2, \ldots, b_q]^\mathrm{T}$ as output. The output $\bm{z}_{br}^{l_{br}+1}$ of the feed-forward branch neural network is expressed as
\begin{equation} \label{eq:output_branch}
    \begin{split}
    \bm{z}_{br}^{l_{br}+1} &= \left[b_1, b_2, \ldots, b_q\right]^\mathrm{T}\\
    &=\mathcal{R}_{br}\left(\bm W^{l_{br}}\bm{z}^{l_{br}} + \bm b^{l_{br}+1}\right),
    \end{split}
\end{equation}
where $\mathcal{R}_{br}\left( \cdot \right)$ denotes the nonlinear activation function for the branch net and $\bm{z}^{l_{br}} = f_{br}(u_i(\bm x_1), u_i(\bm x_2), \ldots, u_i(\bm x_m))$, where $f_{br}\left(\cdot\right)$ denotes a branch net function. Similarly, consider a trunk network with $l_{tr}$ hidden layers, where the $(l_{tr}+1)$-th layer is the output layer consisting of $q$ neurons. The trunk net outputs a feature embedding $[t_1, t_2, \ldots, t_q]^\mathrm{T}$. The output of the trunk network can be represented as
\begin{equation} \label{eq:output_trunk}
    \begin{split}
    \bm{z}_{tr}^{l_{tr}+1} &= \left[t_1, t_2, \ldots, t_q\right]^\mathrm{T}\\
    &=\mathcal{R}_{tr}\left(\bm W^{l_{tr}}\bm{z}^{l_{tr}} + \bm b^{l_{tr}+1}\right),
    \end{split}
\end{equation}
where $\mathcal{R}_{tr}\left( \cdot \right)$ denotes the non-linear activation function for the trunk net and $\bm{z}^{l_{tr}-1} = f_{tr}(\bm y_1, \bm y_2, \ldots, \bm y_p)$. The key point is that we uncover a new operator $\mathcal G_{\bm{\theta}}$ as a neural network that can infer quantities of interest from unseen and noisy inputs. The two networks are trained to learn the solution operator such that
\begin{equation}
    \mathcal G_{\bm{\theta}}:\bm{u}_i \rightarrow \mathcal G_{\bm{\theta}}(\bm{u}_i),\;\; \forall\;\; i = \{1,2,3, \ldots, N\}.
\end{equation}
For a single input function $\bm u_i$, the DeepONet prediction $\mathcal G_{\bm \theta}(\bm u)$ evaluated at any coordinate $\bm y$ can be expressed as
\begin{equation} \label{eq:output_deeponets}
    \begin{split}
    \mathcal G_{\bm{\theta}}(\bm u_{i})(\bm y) &= \sum_{k = 1}^{q}\left(\mathcal{R}_{br}(\bm W^{l_{br}}_k\bm{z}^{l_{br}-1}_k + \bm b^{l_{br}}_k)\cdot \mathcal{R}_{tr}(\bm W^{l_{tr}}_k\bm{z}^{l_{tr}-1}_k + \bm b^{l_{tr}}_k)\right)\\
    &= \sum_{k = 1}^{q}b_k(u_i(\bm x_1), u_i(\bm x_2), \ldots, u_i(\bm x_m))\cdot t_k(\bm y).
    \end{split}
\end{equation} 
DeepONet requires large annotated datasets of paired input-output observations, but it provides a simple and intuitive model architecture that is fast to train, allowing a continuous representation of the target output functions that is resolution-independent. Conventionally, the trainable parameters of the DeepONet represented by $\bm{\theta}$ in \autoref{eq:output_deeponets} are obtained by minimizing a loss function. Common loss functions used in the literature include the $L_1$- and $L_2$-loss functions, defined as
\begin{equation} \label{eq:L1_loss}
\begin{split}
    \mathcal L_1 &= \sum_{i =1}^n \sum_{j =1}^p \big| \mathcal G(\bm u_{i})(\bm y_j) - \mathcal G_{\bm{\theta}}(\bm u_{i})(\bm y_j)\big|\\
    \mathcal L_2 &= \sum_{i =1}^n \sum_{j =1}^p\big(\mathcal G(\bm u_{i}(\bm y_j) - \mathcal G_{\bm{\theta}}(\bm u_{i})(\bm y_j)\big)^2,\\
\end{split}
\end{equation}
where $\mathcal{G}_{\bm{\theta}}(\bm u_{i})(\bm y_j)$ is the predicted value obtained from the DeepONet, and $\mathcal G(\bm u_{i})(\bm y_j)$ is the target value.

Next, we present a DeepONet algorithm for the diesel engine where we compute the weights and biases associated with the deep neural networks based on the available labelled datasets. 

\subsection{Surrogate DeepONet model for Diesel Engine}
\label{subsec:surrogate_model}

\begin{figure}[!h]
  \centerline{\includegraphics[width = 1\textwidth]{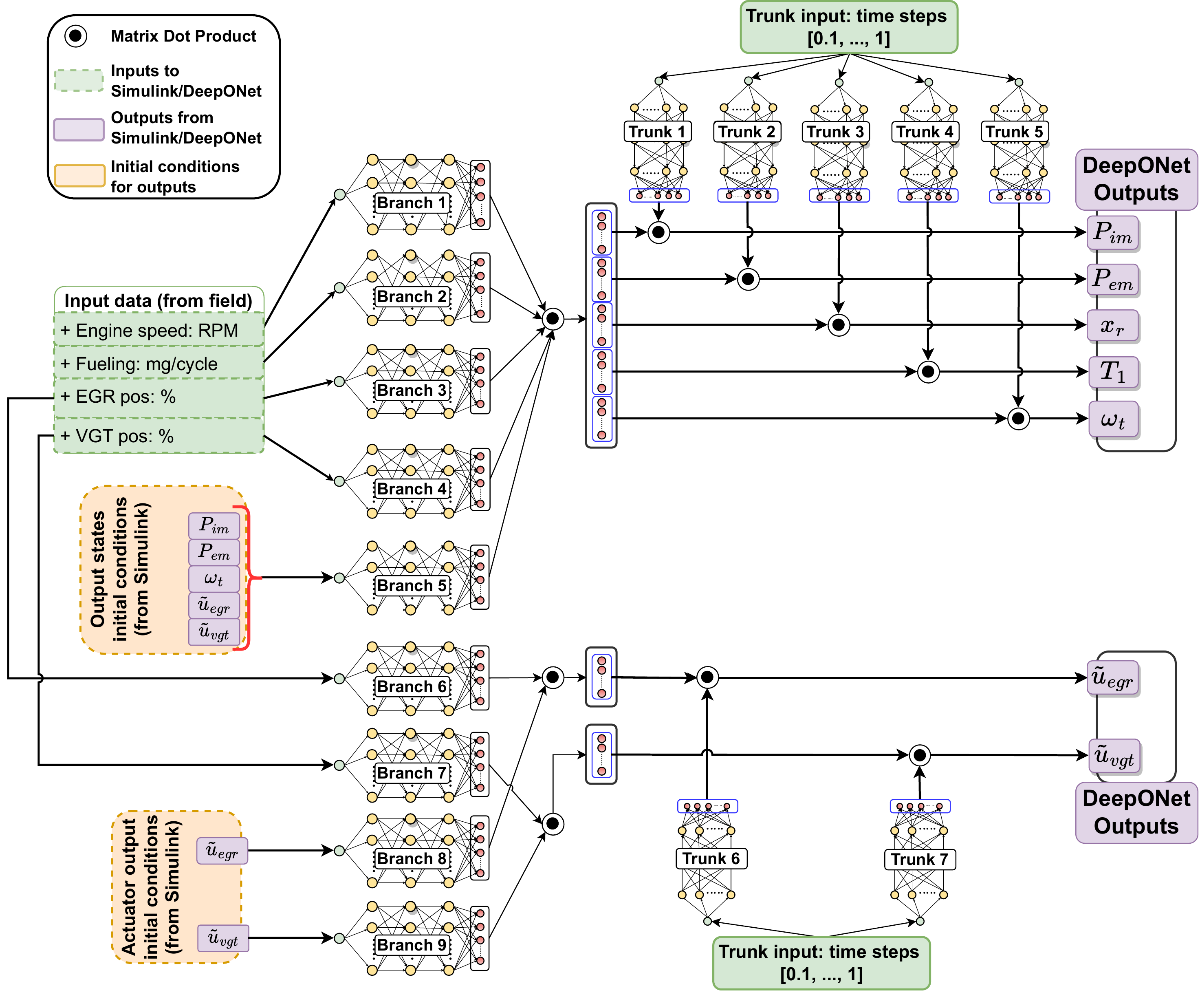}}
  \caption{Multi-input DeepONet architecture to approximate the output states of the Diesel engine. The inputs to the model are encapsulated in green boxes, while the outputs are in purple coloured boxes. The architecture employs nine branch networks for generating the functional mapping whereas the seven trunk networks are used for determining the basis functions (at each temporal point) for each output states. Separate branches are used for $\tilde u_{egr}$ and $\tilde u_{vgt}$ due to their strict dependence on $u_{egr}$ and $u_{vgt}$. Initial conditions for the output states are provided as additional inputs to assist in solution convergence.}
\label{fig:DeepONet_schematic}
\end{figure}

Conventionally, the neural operators are designed to take a single function as an input in the branch network. 
However, in designing a surrogate model for the diesel engine, we have to consider four input functions, $n_e, u_{\delta}, u_{egr}, u_{vgt}$. The limitation of the input space in the conventional DeepONet architecture prohibits us to learn a wide range of useful operators defined on multiple input spaces. To that end, we employ a multiple-input operator architecture of DeepONet \cite{jin2022mionet, goswami2022neural} in our proposed surrogate model. The proposed architecture used in this work to approximate the output states of the diesel engine is shown in figure \ref{fig:DeepONet_schematic}.

\begin{figure}[h]
  \centering
  \centerline{\includegraphics[width = 0.6\textwidth]{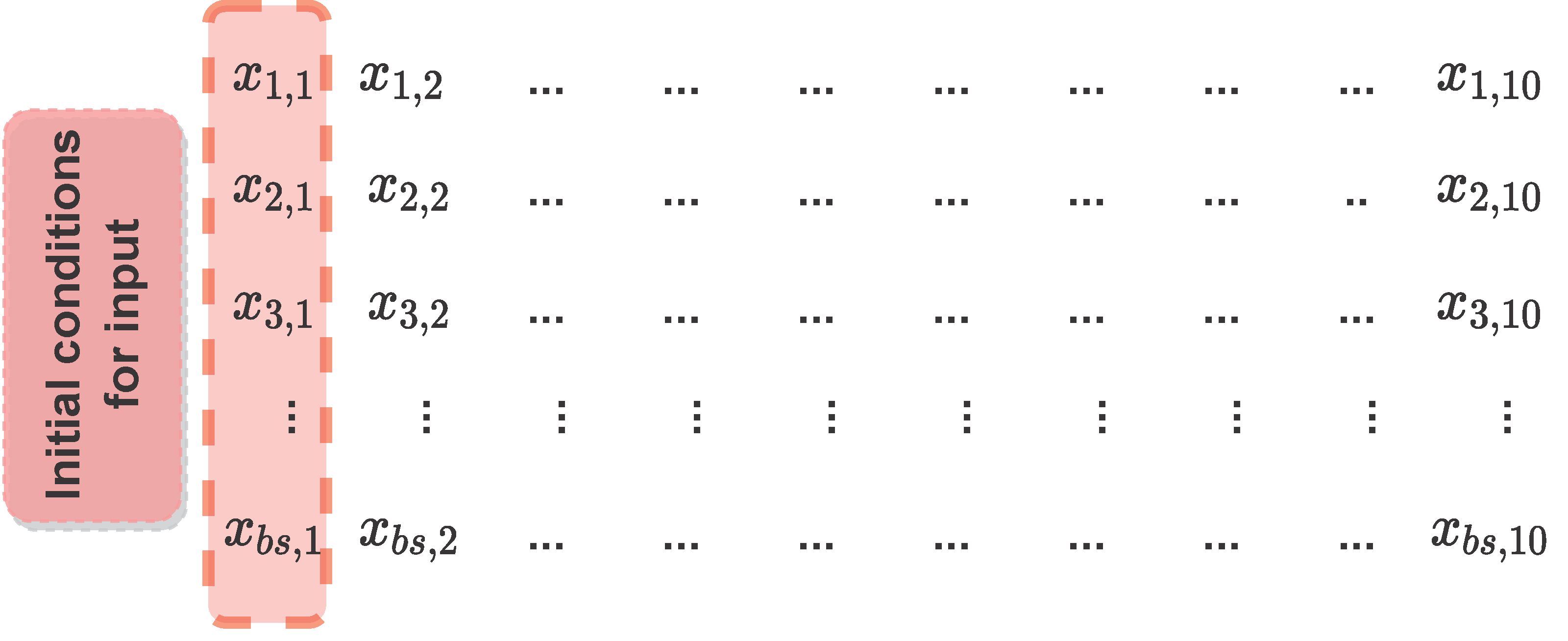}}
  \caption{Extracting initial conditions for use as input to DeepONet from output field. These initial conditions are only provided for measurable outputs $P_{im}, P_{em}, \omega_{t}, \tilde u_{egr},$ and $\tilde u_{vgt}$.}
\label{fig:DeepONet_init_conds}
\end{figure}

To prepare the training data for DeepONet, we divide the temporal signals of the four inputs, $n_e, u_{\delta}, u_{egr}, u_{vgt}$, collected from engine test bed in steps of ten time points per signal, thereby converting a point based signal into a feature-based representation to enhance the network's learning process. Each of the four input signals are associated with a branch network (Branch$1-4$ in figure \ref{fig:DeepONet_schematic}) to learn features that works as a dedicated function approximator. In addition to the four inputs, another set of inputs is provided to the DeepONet in the form of initial conditions, extracted from the output predictions in Branch$5$.  This is analogous to providing initial conditions for solving differential equations. Here, the initial conditions are extracted from the first input point for each training signal. See figure \ref{fig:DeepONet_init_conds} showing the data extraction process for initial conditions from outputs. Note that the initial condition values are only used for output states that can be measured in field, namely $P_{im}$, $P_{em}$, $\omega_{t}$, $\tilde u_{egr}$, and $\tilde u_{vgt}$ (see figure \ref{fig:engine_schematic}). Providing initial conditions as inputs helps in bounding the operator learning process, analogous to the process of solving differential equations. The dot product of the output embeddings of branch networks $1-5$ and trunk networks $1-5$ maps to the five output states, $P_{im}$, $P_{em}$, $x_{r}$, $T_{1}$, and $\omega_{t}$. The functional representation for the inputs, $u_egr$ and $u_vgt$ is generated separately for branch networks $6$ and $7$, which when coupled with the branch networks $8$ and $9$, that contains information about the initial conditions of $\tilde u_{egr}$ and $\tilde u_{vgt}$, and the corresponding output of the trunk networks $6$ and $7$ approximates the output states, $\tilde u_{egr}$ and $\tilde u_{vgt}$. Separate branches were used for generating states $\tilde u_{egr}$ and $\tilde u_{vgt}$ since out of the four inputs, only $u_{egr}$ and $u_{vgt}$ are associated with them. Interested readers can find more details from Equations $39-41$ and $57$ in \cite{wahlstrom2011modelling}. Understanding the associativity between output states and the inputs that affect them is important since it assists the network learning process and drawing appropriate mapping between inputs and outputs. Here, we emphasize that this proposed design can be altered as needed to incorporate additional inputs owing to the DeepONet architecture's flexibility. For instance, the design can be changed to incorporate an extra branch and trunk if the network needs to be expanded to include a new input and/or output state. We use the associativity of pertinent equations described in \cite{wahlstrom2011modelling} to map appropriate inputs to output states.

\subsection{Data generation}
\label{subsec:data_gene}

\begin{figure}[h]
  \centering
  \centerline{\includegraphics[width = 1\textwidth]{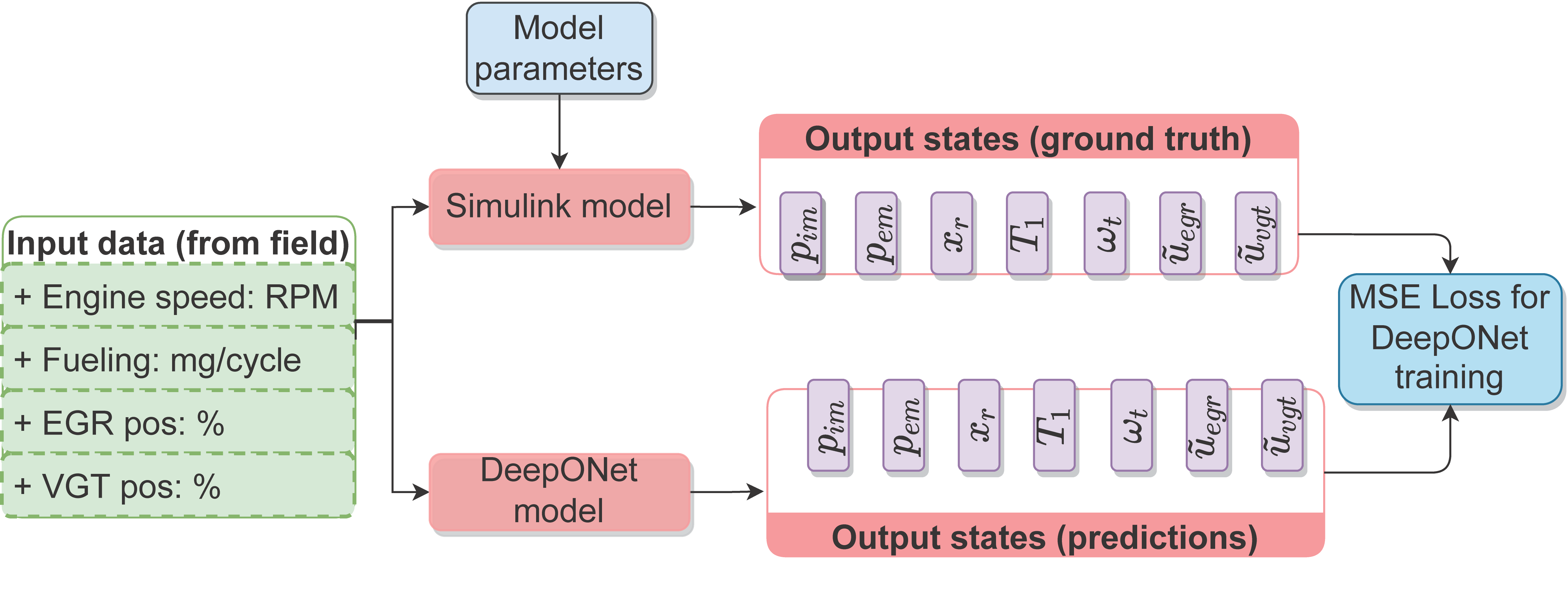}}
  \caption{Overview of DeepONet model and its use in conjunction with Simulink model as ground truth generator.}
\label{fig:model_overview}
\end{figure}

In order to train the deep neural networks, the ground truth for the output states is generated using the Simulink model as discussed in Section \ref{sec:numerical_simulation}. Figure \ref{fig:model_overview} shows the proposed scheme for training the surrogate model using data generated from Simulink. The mean squared error between the DeepONet prediction and Simulink's output is used for back-propagation and network training. In this work, the Simulink model is simulated using the existing parameter values obtained from \cite{wahlstrom2011modelling}.

As discussed previously, the initial conditions are obtained from the labelled output datasets. However, the initial conditions of the output states may not be available in the real setup or one may want to compare the output from measured sensors with expected output emanating from our surrogate model to identify discrepancy in system behavior. To overcome this challenge, we pose the problem a \emph{sequence-to-sequence} learning. The initial conditions of the first test signal, $\mathcal S_{\text{test}}^1$, is obtained from the output of the last temporal point of the last training signal. Subsequently, the $t$-th test signal, $\mathcal S_{\text{test}}^t$ fetches the initial condition from the predicted output of the last temporal point of the signal, $\mathcal S_{\text{test}}^{(t-1)}$, obtained as a prediction using the proposed surrogate model.

\emph{Data partition:} The input and ground truth datasets were divided into training and testing sets by choosing a continuous chunk of time. The total time duration for the complete dataset was approximately 15 hours and was collected over a period of time with the same engine. From this, a continuous section of 1000 sec was allocated for use as testing data. The testing data segment was chosen carefully to ensure its representation in the training set. Less frequently occurring scenarios such as engine idling were not considered in testing since the dataset lacked sufficient idling condition data points for the model to learn. It should be noted that during input data generation, all possible engine conditions that may be encountered need to be taken into account. Using a rich input space for training enables the DeepONet network to make generalized predictions. The input data is first transformed into signal trains with a window size of $10$ (corresponding to a 5 second data chunk). The objective of this transformation is to provide the neural network model with larger features to enhance learning. The window size is chosen heuristically through experimentation with different window size. It was observed that the model accuracy reduces as the window size is made larger. All four inputs are then converted into signal trains before using as inputs for the DeepONet. 

\section{Experimental results}
\label{sec:results}

In this section, we present and discuss the results from our experiments. In addition to experiments with clean input data, we also present results with noisy inputs and outputs to show the sensitivity of DeepONet predictions for similar problems. 

\subsection{DeepONet prediction results}
\label{subsec:deeponet_results}

The architecture details of the surrogate DeepONet model shown in figure  \ref{fig:DeepONet_schematic} are shown in table \ref{Tab:arch_details}. For weight optimization, the Adam optimizer with learning rate scheduler is used. The starting learning rate used was $1e-3$ until first $5,000$ epochs, which was then reduced to $5e-4$ until 10,000 epochs. Thereafter, a constant learning rate of $1e-4$ was used until the training terminates at $20,000$ epochs. Dropout was also incorporated as a network regularizer in the branch networks that generate the functional representation of the inputs. Dropout rates were tuned heuristically. In addition to regularizing the network, dropout can also be used for model uncertainty estimation \cite{gal2016dropout}, and we provide uncertainty estimation results in the following sections. The model was trained on a NVIDIA A40 GPU and the training time was approximately three hours. Point-wise self-adaptive weights were also used to regularize each temporal point during training \cite{mcclenny2020self,kontolati2022influence}. The resulting $\mathcal L_2$ error values for the seven output states over a prediction time window of $1000$ seconds is shown in table \ref{Tab:Error_noise} (first row).

\begin{table}[H]
\centering
\caption{Details of DeepONet architecture used for diesel engine modeling}
{\begin{tabular}{cccccc} 
\toprule
\multirow{2}{*}{\textbf{Network}}  & \multirow{2}{*}{\textbf{Depth}}  & \multirow{2}{*}{\textbf{Width}} & \multicolumn{2}{c}{\textbf{Activation function}} & \multirow{2}{*}{\textbf{Dropout Rate}}\\  \cmidrule(l){4-5} & & & \textbf{Hidden layers} & \textbf{Final layer} & \\
\hline
Branch $1-2$ & $3$ & $128$ & \texttt{Swish} & \texttt{ReLU} & $0.20$ \\
Branch $3-4$ & $3$ & $128$ & \texttt{Swish} & \texttt{ReLU} & $0.10$ \\
Branch $6-7$ & $2$ & $128$ & \texttt{Swish} & \texttt{Linear} & $0.05$ \\
Branch $8-9$ & $3$ & $128$ & \texttt{Swish} & \texttt{Linear} & $--$ \\
Trunks & $3$ & $128$ & \texttt{Swish} & \texttt{Sigmoid} & $--$ \\
\bottomrule
\end{tabular}}
\label{Tab:arch_details}
\end{table}

\begin{figure}[!h]
  \centering
  \centerline{\includegraphics[width = 1\textwidth]{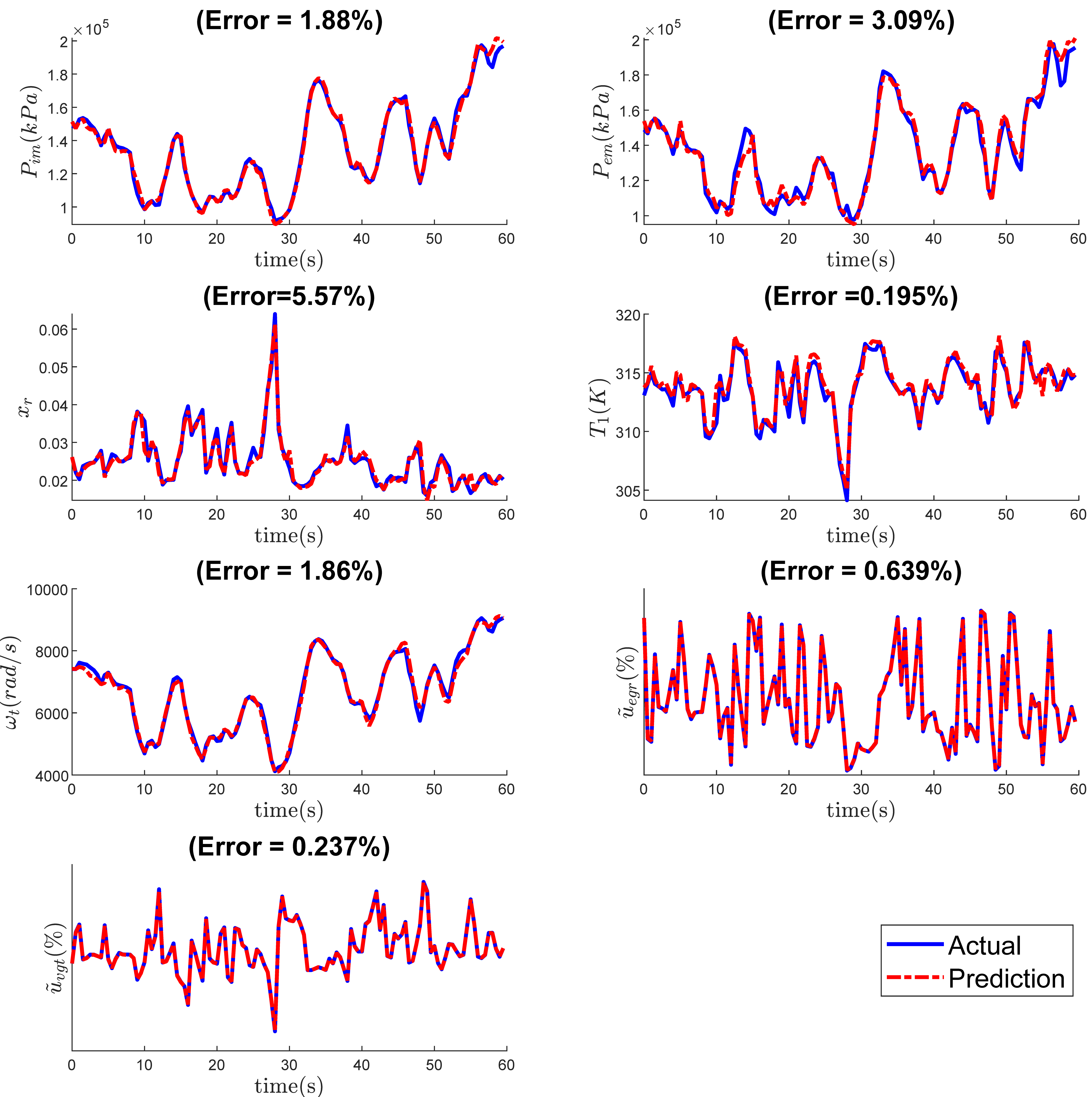}}
  \caption{DeepONet prediction results for the seven output states and comparison with ground truth obtained from Simulink. The time subset is limited to the first $60$ seconds here for clarity. Error value labels on top of each plot are values evaluated for this 60 second time window. Y-axis markers for $\tilde u_{egr}$ and $\tilde u_{vgt}$ are not shown in interest of confidentiality.}
\label{fig:DeepONet_prediction}
\end{figure}

\begin{figure}[!h]
  \centering
  \centerline{\includegraphics[width = 0.6\textwidth]{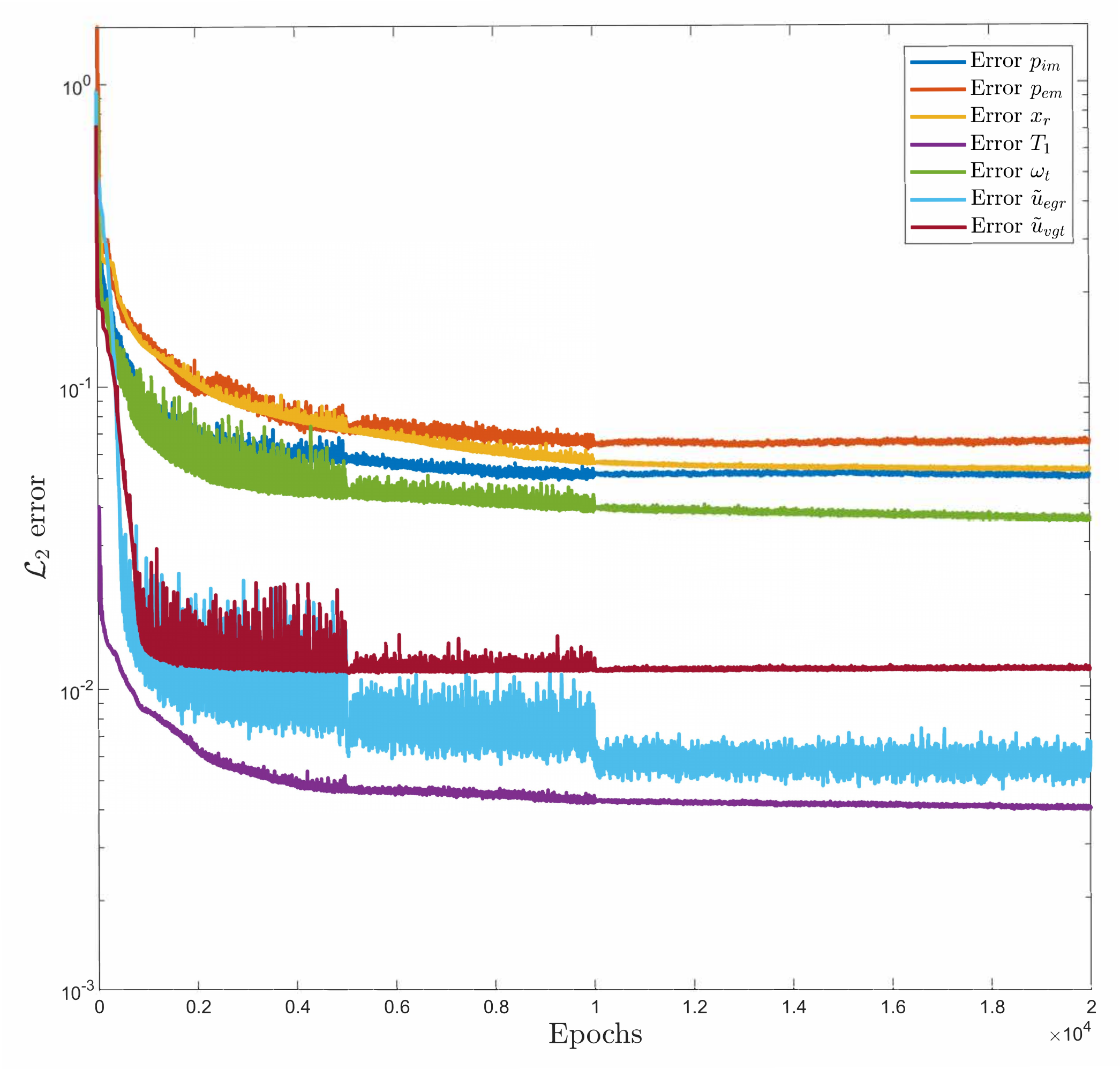}}
  \caption{Test error propagation for the seven output states during the training. The different levels in the error propagation for each output which is likely due to the difference in operator mapping complexity for each function.}
\label{fig:DeepONet_error}
\end{figure}

As a representative case, figure \ref{fig:DeepONet_prediction} shows the comparison between DeepONet predictions and ground truth for the first $60$ second time window in the testing data subset. The output state $P_{em}$ has the highest error rate of $6.5\%$ for the $1000$ sec evaluation window. Predictions for actuator signal states $\tilde u_{egr}$ and $\tilde u_{vgt}$ show a good match to their ground truth values. This is expected since their functional relationships is easier to map as per their driving ODEs. $P_{em}$ represents the pressure in the exhaust manifold and the combustion process inside the cylinder. State $x_{r}$, which indicates residual gas fraction inside the combustion chamber, is an important parameter in determining the state of exhaust gases, especially its temperature and thereby pressure. In our architecture, the initial conditions for $x_{r}$ were not used as input since this state cannot be measured directly in the field. The error in $x_{r}$ estimation could likely have a strong impact on the error for state $P_{em}$ since the network is trained together, leading to higher prediction error for $P_{em}$. The higher error for $P_{em}$ could also indicate the need to identify additional input features to help augment operator learning process for this state. Figure \ref{fig:DeepONet_error} shows the error propagation for the seven states during model training.

\subsection{Results with noisy data}
\label{subsec:results_noisyInputs}

As with any real systems, diesel engine sensor signals are associated with noise that is difficult to ascertain or determine with certainty. To understand the impact of noisy input on model prediction, we simulate three noise conditions: $1\%$, $2\%$, and $3\%$ additive white Gaussian noise. The level of noise chosen is based on the expected noise levels from sensors generating the four inputs to the DeepONet on a real system. This white noise is added to the input data and the prediction is made using proposed architecture of DeepONet that has been pre-trained on clean data as discussed earlier . We test with noisy data on the same section in time as the validation dataset with clean data. This ensures that prediction is made on chunk of data not seen by the network during its training.

\begin{figure}[h!]
  \centering
  \centerline{\includegraphics[width = 1\textwidth]{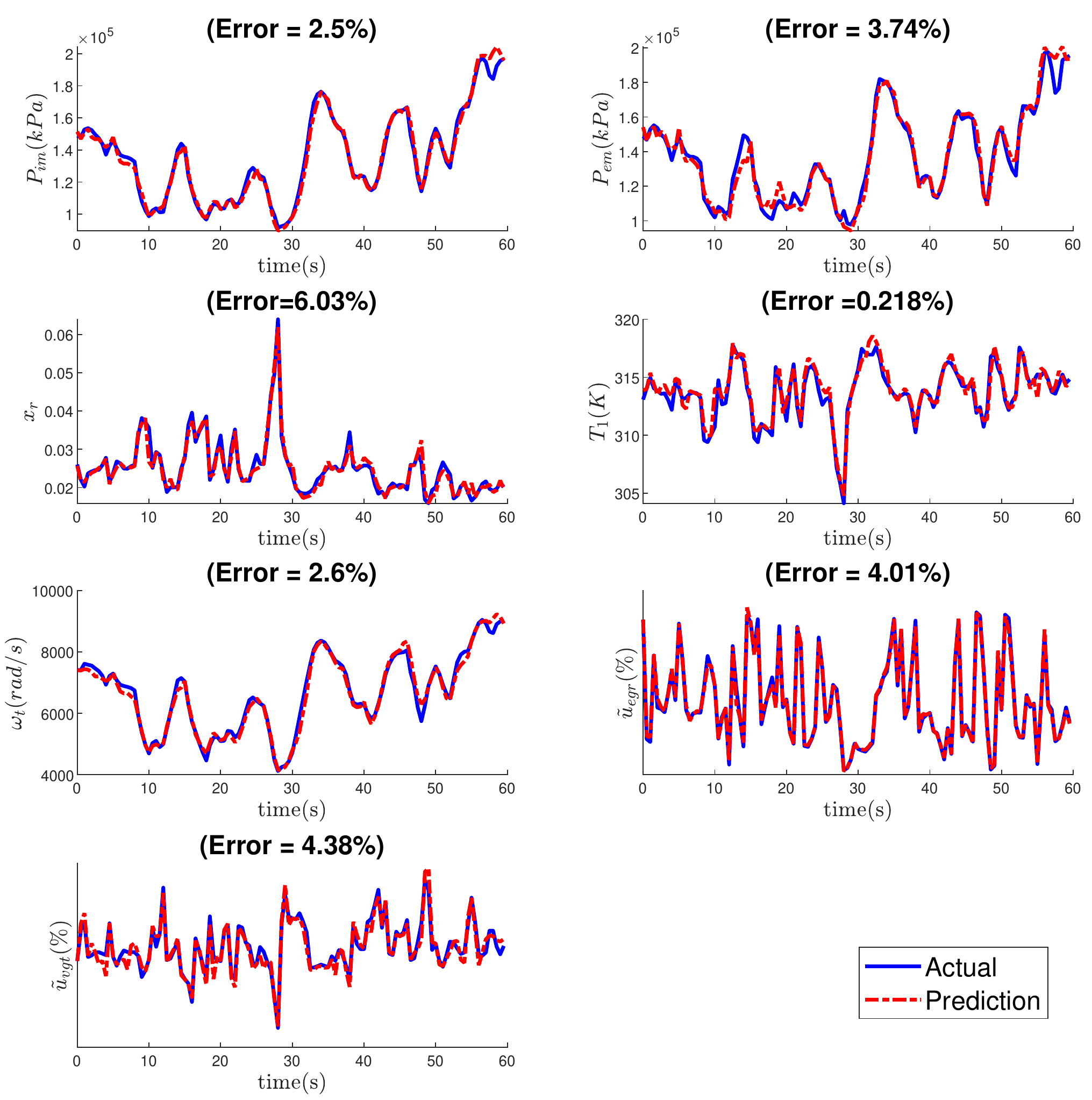}}
  \caption{DeepONet prediction results with $3\%$ Gaussian white noise added to testing input data subset. Prediction error increases progressively with noise level with the actuator predictions showing highest increase possibly due to the low dropout rate used for their respective branch networks.}
\label{fig:Predictions_noise}
\end{figure}

Table \ref{Tab:Error_noise} shows the error comparison between prediction results with clean inputs vs Gaussian white noise in input testing dataset. The two output states, $\tilde u_{egr}$ and $\tilde u_{vgt}$ are most affected by the addition of input noise. The effect of noise addition on other output states is minimal. Figure \ref{fig:Predictions_noise} shows a comparison between ground truth and DeepONet prediction with a $3\%$ Gaussian white noise associated with input testing data. 

\begin{table}[!h]
\centering
\caption{$\mathcal L_2$ error values for the predicted seven output states obtained using the proposed DeepONet based surrogate model. The error results are over a testing window of $1000$ seconds. Comparison of error for no noise and different noise conditions is also shown here.}
{%
\begin{tabular}{cccccccc} 
\toprule
\multicolumn{1}{c}{\multirow{2}{*}{\textbf{Noise~level }}} & \multicolumn{7}{c}{\textbf{$\mathcal{L}_2$ error (\%)}} \\ 
\cmidrule{2-8}
\multicolumn{1}{c}{} & \textbf{\textit{$P_{im}$}} & \textbf{\textit{$P_{em}$}} & \textbf{\textit{$x_{r}$}} & \textbf{\textit{$T_{1}$}} & \textbf{\textit{$\omega_{T}$}} & \textbf{\textit{$\tilde u_{egr}$}} & \textbf{\textit{$\tilde u_{vgt}$}} \\
\hline
No noise & $4.96$ & $6.44$ & $5.22$ & $0.39$ & $3.56$ & $0.59$ & $1.15$ \\
$1\%$ & $5.09$ & $6.64$ & $5.26$ & $0.40$ & $3.66$ & $1.54$ & $1.86$ \\
$2\%$ & $5.15$ & $6.75$ & $5.30$ & $0.41$ & $3.71$ & $2.64$ & $3.03$ \\
$3\%$ & $5.17$ & $6.94$ & $5.56$ & $0.41$ & $3.89$ & $4.14$ & $4.35$ \\
\bottomrule
\end{tabular}}
\label{Tab:Error_noise}
\end{table}

We also evaluated the sensitivity of our model prediction when subjected to noise in output data. The objective of this study was to understand the impact of noise on model's accuracy when model is trained on dataset which is inherently noisy, as is the case for most field measurements. For this, we add Gaussian white noise of magnitude up to $3\%$ to the output training dataset. The $3\%$ limit on noise level was chosen based on field experience with sensors that measure the output states in this study. The DeepONet is trained with this noisy labeled data, and the trained network is then used for generating prediction with clean input and output dataset. Table \ref{Tab:Error_noise_output} shows the error comparison between the DeepONet model trained with clean vs noisy outputs. 

\begin{table}[!h]
\centering
\caption{$\mathcal L_2$ error comparison between model trained with clean vs noisy output training dataset. Marginal increase in relative error is observed for the network trained with added noise to labeled training data. Predictions here were made using noise-free test dataset.}
{%
\begin{tabular}{cccccccc} 
\toprule
\multicolumn{1}{c}{\multirow{2}{*}{\textbf{Noise~level }}} & \multicolumn{7}{c}{\textbf{$\mathcal{L}_2$ error (\%)}} \\ 
\cmidrule{2-8}
\multicolumn{1}{c}{} & \textbf{\textit{$P_{im}$}} & \textbf{\textit{$P_{em}$}} & \textbf{\textit{$x_{r}$}} & \textbf{\textit{$T_{1}$}} & \textbf{\textit{$\omega_{T}$}} & \textbf{\textit{$\tilde u_{egr}$}} & \textbf{\textit{$\tilde u_{vgt}$}} \\
\hline
No noise & $4.96$ & $6.44$ & $5.22$ & $0.39$ & $3.56$ & $0.59$ & $1.15$ \\
$3\%$ & $4.97$ & $7.14$ & $5.47$ & $0.85$ & $5.17$ & $0.67$ & $1.17$ \\
\bottomrule
\end{tabular}}
\label{Tab:Error_noise_output}
\end{table}

\subsection{Sequence-to-sequence chaining of initial conditions}
\begin{figure}[H]
  \centering
  \centerline{\includegraphics[width = 0.75\textwidth]{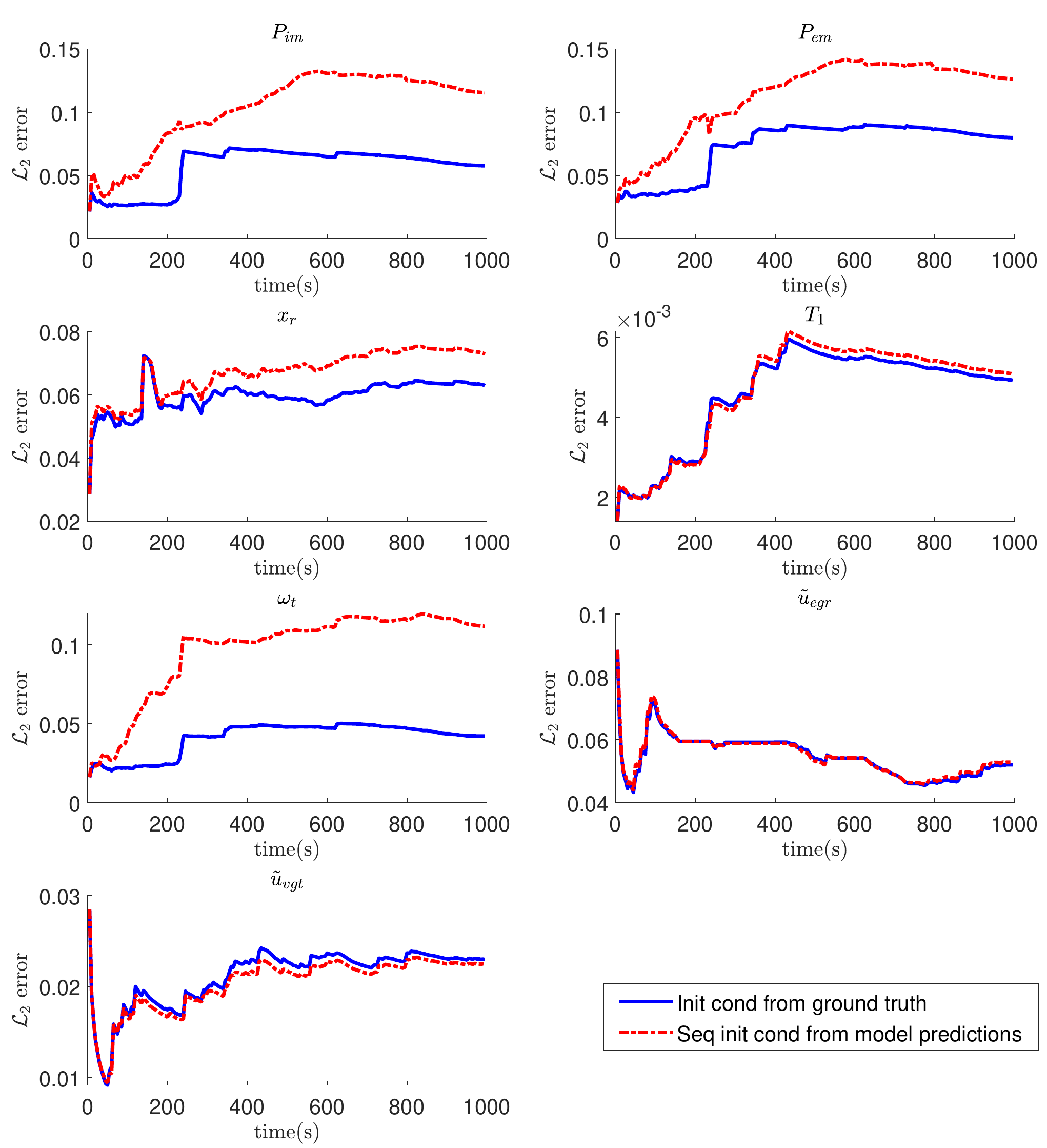}}
  \caption{Comparison of cumulative error between using sequence-to-sequence scheme for initial condition vs using initial conditions for signal trains from ground truth. A higher error accumulation is seen when the sequence-to-sequence scheme is used due to the prediction inaccuracy that results from DeepONet model prediction when used as the initial condition for next signal train. This cumulative error is highest for states $P_{im}$, $P_{em}$, and $\omega_{t}$ possibly due to the higher error values for the DeepONet model itself.}
\label{fig:error_prop_seq_to_seq}
\end{figure}

In addition to using initial conditions from the ground truth output data (generated from Simulink), we evaluated the effect of sequence-to-sequence chaining of initial conditions for our data train signals. This is important in real applications where the ground truth data is not available or maybe erroneous due to systematic faults with the data logging system. Here, we take the predicted DeepONet output for the previous signal train and use it as an initial condition for the next signal train. This sequence is followed for all the testing dataset and the complete prediction is obtained by processing one signal train at a time. This, however leads to longer prediction time ($\approx$ 30 seconds in our case) and higher prediction error accumulation as the prediction sequence gets longer. Figure \ref{fig:error_prop_seq_to_seq} shows the comparison between cumulative error between sequence-to-sequence initial condition use scheme vs using initial conditions from ground truth data. Table \ref{Tab:seq_seq_error} shows a comparison between relative error during prediction for the two initial condition schemes.

\begin{table}[!h]
\centering
\caption{$\mathcal L_2$ error comparison between prediction with ground truth as initial condition vs sequence to sequence initial condition generation scheme. }
{%
\begin{tabular}{cccccccc} 
\toprule
\multicolumn{1}{c}{\multirow{2}{*}{\textbf{Initial condition from }}} & \multicolumn{7}{c}{\textbf{$\mathcal{L}_2$ error (\%)}} \\ 
\cmidrule{2-8}
\multicolumn{1}{c}{} & \textbf{\textit{$P_{im}$}} & \textbf{\textit{$P_{em}$}} & \textbf{\textit{$x_{r}$}} & \textbf{\textit{$T_{1}$}} & \textbf{\textit{$\omega_{T}$}} & \textbf{\textit{$\tilde u_{egr}$}} & \textbf{\textit{$\tilde u_{vgt}$}} \\
\hline
Ground truth & $4.96$ & $6.44$ & $5.22$ & $0.39$ & $3.56$ & $0.59$ & $1.15$ \\
Seq-to-seq & $11.54$ & $12.64$ & $7.28$ & $0.85$ & $11.17$ & $5.29$ & $2.25$ \\
\bottomrule
\end{tabular}}
\label{Tab:seq_seq_error}
\end{table}

\section{Model uncertainty estimation}
\label{sec:uncertainty_eval}
Neural networks, when trained to learn a certain set of weights and biases to predict outputs are deterministic, and the same inputs generate the same outputs. Model uncertainty pertains to the potential variations that may exist in the model's weight estimates and understanding the amount of uncertainty can help in building confidence around the model's prediction. Probabilistic modeling of neural networks has been used as an uncertainty estimation tool in deep learning research \cite{mackay1992practical,jospin2022hands,psaros2022uncertainty}. These methods are rooted in the Bayesian probabilistic framework. Traditional approaches for Bayesian modeling include Monte Carlo Dropout \cite{gal2016dropout,srivastava2014dropout}, HMC \cite{hastings1970monte, bardenet2017markov, neal2011mcmc}, Bayes-by-backprop \cite{blundell2015weight, hernandez2015probabilistic, blei2017variational} and others; see a comprehesive review in \cite{psaros2022uncertainty}. In this work, we use dropout to understand model uncertainty due to the following reasons:
\begin{itemize}
    \item  Dropout is as regularizer in the DeepONet Branch layers.
    \item Dropout provides a faster approach for model uncertainty estimation. Other approaches such as Variational Inference can be challenging to train on complex architectures such as ours. Additionally, setting up prior and posterior distribution \cite{meng2022learning} presents additional complexities with our sensory data.
\end{itemize}

\noindent We demonstrate model uncertainty through the following steps:
\begin{enumerate}
    \item  Train the DeepONet model with the input data (No noise) using MC Dropout in branch networks, with the architecture presented in table \ref{Tab:arch_details}. 
    \item Generate predictions on test section of the data. Each time a prediction is made, a different prediction of the output states is obtained due to the stochastic nature imparted to branch networks by the dropouts layers.
    \item Collect $100$ outputs and generate an ensemble mean to represent the mean prediction, $\mu$. Calculate the standard deviation, $\sigma$ for each point for all predictions thus generated.
    \item Create an uncertainty region around the ensemble mean with a spread of $\mu \pm 2\sigma$
\end{enumerate}

\begin{figure}[H]
  \centering
  \centerline{\includegraphics[width = 0.90\textwidth]{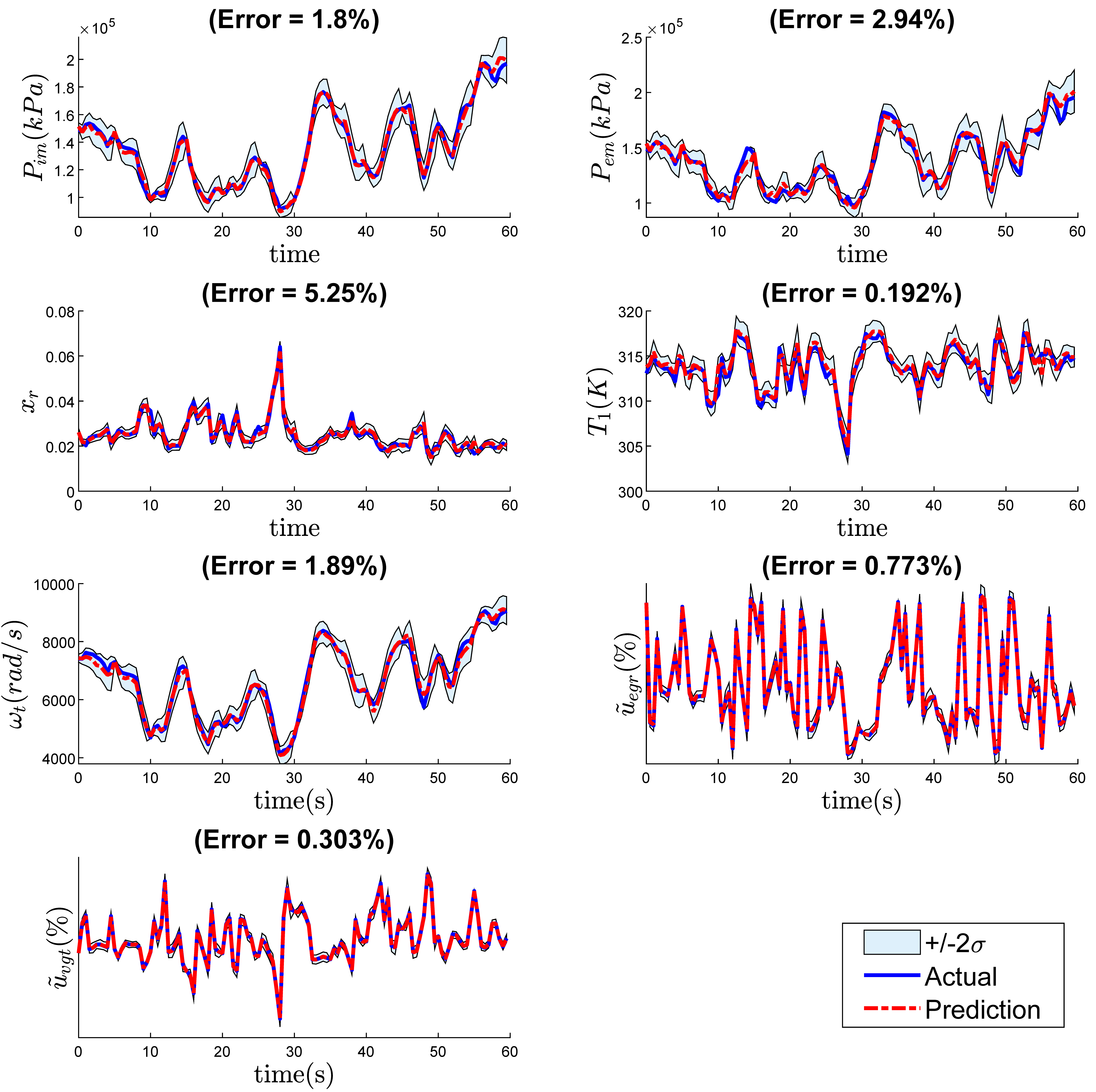}}
  \caption{DeepONet prediction and uncertainty results with an ensemble mean generated by drawing 100 samples during prediction in a MC dropout network with dropout rates as per \ref{Tab:arch_details}. The predicted value is the mean of $100$ samples while the uncertainty, represented by the standard deviation for these samples is shown as a $\pm 2\sigma$ zone around the ensembled mean.}
\label{fig:Predictions_ensm}
\end{figure}

Figure \ref{fig:Predictions_ensm} shows the comparison between ensemble mean that is calculated by generating 100 predictions from the Dropout based DeepONet model. The standard deviation for each point is calculated over these $100$ predictions to generate the uncertainty band around this ensembled mean. Table \ref{tab:ensemble_error} shows the total error for $1000$ seconds testing subset with an ensemble mean of 100 predictions from a dropout-based stochastic DeepONet. Relative $\mathcal L_2$ error \% for the seven output states estimated using the ensemble mean have higher error as compared to the deterministic prediction errors. This is due to the stochastic nature of the branch layers that use Dropout that results in the network learning different mapping each time a prediction is made. To demonstrate the maximum prediction error possible with this model, we calculate the worst-case prediction boundary by using the $\pm 2\sigma$ value with ensemble mean and estimate the error with respect to the target prediction. The worst-case error values are also shown in table \ref{tab:ensemble_error}. The maximum error possible in prediction estimates in this worst case scenario is approximately $12.6\%$ for state $P_{em}$. 

\section{Limitations}
\label{sec:limitations}

\begin{figure}[h]
  \centering
  \captionsetup{justification=centering,margin=0.1cm}
  \centerline{\includegraphics[width = 0.9\textwidth]{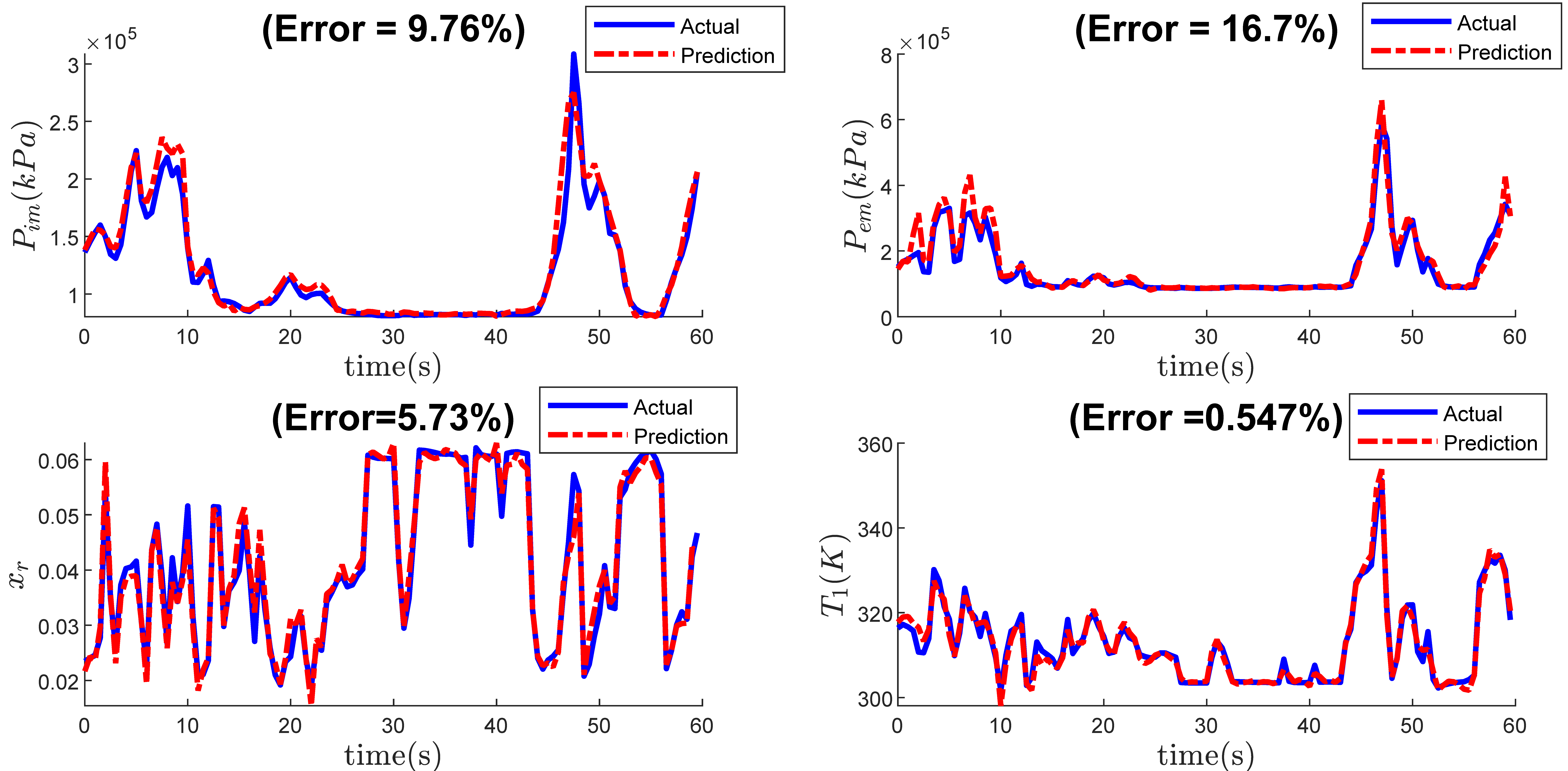}}
  \caption{DeepONet predictions for a different time segment from the testing dataset showing limitations for the current model with the training dataset used in this work. High error \% are observed in this case possibly due to lack of representative data points in the training dataset.}
\label{fig:Predictions_variation1}
\end{figure}

In this section, we discuss some of the limitations of the proposed operator network-based surrogate model. The error values for the predictions discussed in earlier sections (Table \ref{Tab:Error_noise} and \ref{tab:ensemble_error}) are the total error values calculated across the $1000$ second subset used in testing. However, the error value is not uniform across this entire time span, and varying results are seen for different time segments. Figure \ref{fig:Predictions_variation1} shows the prediction results for the next 60 second time window compared to what was shown in figure \ref{fig:DeepONet_prediction} from the testing dataset. Here, we notice significantly higher error values for the four output states when compared to results shown in figure \ref{fig:DeepONet_prediction}. This

\begin{table}[h]
\centering
\caption{Error comparison between deterministic prediction of the output states from DeepONet with an ensemble of predictions generated by using Dropouts in the branch network. Error values for ensemble mean indicates the total error for the $1000$ second time sequence that was used for testing when $100$ different predictions were ensembled. Error value for ensemble $\mu + 2\sigma$ indicates the worst-case error when the prediction is at the boundary of the uncertainty band.}
{%
\begin{tabular}{lccccccc} 
\toprule
\multirow{2}{*}{\textbf{Testing condition}} & \multicolumn{5}{c}{\textbf{$\mathcal{L}_2$ error (\%)}} \\ \cmidrule{2-8}
  & \textbf{$P_{im}$} & \textbf{$P_{em}$} & \textbf{$x_{r}$} & \textbf{$T_{1}$} & \textbf{$\omega_{t}$} &  \textbf{$\tilde u_{egr}$} & \textbf{$\tilde u_{vgt}$}\\
  \hline
Deterministic prediction & $4.96$ & $6.44$ & $5.22$ & $0.39$ & $3.56$ & $0.59$ & $1.15$ \\
Ensemble $\mu$ & $4.87$ & $6.34$ & $5.22$ & $0.39$ & $3.58$ & $0.77$ & $1.15$ \\
Ensemble $\mu + 2\sigma$ & $8.01$ & $12.60$ & $10.90$ & $0.75$ & $6.81$ & $9.06$ & $3.90$ \\
\bottomrule
\end{tabular}
}
\label{tab:ensemble_error}
\end{table}

observation may be attributed to input data, which was collected from a test bed under generic test conditions. It is possible that the training dataset used here may be limited to predicting only certain operating conditions. Standardizing the training dataset that contains representative information of all possible operating states for the engine should be ideally used for training and evaluation.

Extensions of the current DeepONet architecture to other engine models is possible. However, this would require one to train the DeepONet model using data collected from each individual engine, which involves significant effort and time. We also acknowledge the fact that training a new model where ground truth data is collected from sensory measurements in field might be a challenging task and may require expanding the input features to enable operator learning. This is a research avenue that we continue to explore as a part of our future goal and objective for this work.\\
The current DeepONet model has been trained using the parameter set provided by \citep{wahlstrom2011modelling}. The parameter values required for using the mean-value model for a test engine needs to be determined for data generation. Parameter identification can be accomplished by using the PINN framework as described in \citep{raissi2019physics} and we continue to explore the possibility of combining our operator network with PINN-based approach in future. 

\section{Summary}
\label{sec:summary}
 In this work, we present a deep operator based neural network model for predicting the output states of a mean-value gas flow model for diesel engine. Four input signals, engine speed ($n_{e}$), fueling ($u_{\delta}$), EGR valve position ($u_{egr}$), and VGT valve position ($u_{vgt}$) are used for generating a continuous operator mapping to seven output states. The ground truth for the output states is generated using the Simulink model available online from \cite{Simulink}. The DeepONet model once trained is used for predicting the seven output states using previously unseen testing data. The accuracy and robustness of the DeepONet model are evaluated using $\mathcal{L}_{2}$ relative error with respect to ground truth and accuracy under noisy conditions respectively. We summarize our findings as follows:

\begin{enumerate}
    \item Deep Operator-based neural network model enables learning of operators to convert inputs to desired outputs from a mean-value gas flow model for a diesel engine. The output states predicted by DeepONet show good accuracy over a 1000 sec testing window (see table \ref{Tab:Error_noise}). The maximum relative error observed was $6.4\%$ for output state $P_{em}$. This state has a strong dependence on the exhaust manifold dynamics such as exhaust manifold temperature, and hence it may require additional input features to improve the accuracy. We also observe from figure \ref{fig:DeepONet_prediction} that the states $\omega_{t}$, $P_{im}$, and $P_{em}$ are correlated based on their characteristic shape. This is a consequence of the turbocharger assembly's coupling with the intake and exhaust systems.
    \item DeepONet exhibits good generalization accuracy when tested with simulated noisy data in inputs as well as outputs. Addition of $3\%$ white noise to the inputs during testing on a model trained with noise-free input and labelled data results in a marginal rise in prediction error (see table \ref{Tab:Error_noise}). In addition to adding noise to inputs, we also investigated the consequence of noisy labelled data by training the network with simulated noisy labels and then testing with noise-free data. We observe a marginal rise in prediction error in this experiment as well (see table \ref{Tab:Error_noise_output}). The low sensitivity of our network model to noise may be attributed to the use of Dropouts in our network architecture which enables the neural network to learn more generalized solutions for regression tasks. 
    \item To enable the use of our architecture in real scenarios where the ground truth may not be available or trustworthy, sequence-to-sequence linking of initial conditions is proposed. The last output prediction from the current signal train is used as an initial condition to be used as prediction for the next signal train. This process, however leads to higher cumulative error as compared to the situation where the ground truth from output states (when available) is used as initial condition for the model (see table \ref{Tab:seq_seq_error}). From figure \ref{fig:error_prop_seq_to_seq}, we observe that the cumulative error varies with the prediction window which is indicative of the fact that the accuracy of the model varies based on the location of prediction window. We discuss this in more detail in the limitation section \ref{sec:limitations}.
    \item We determine the uncertainty of our DeepONet model through the use of an ensemble mean approach with Dropout layers. The worst case relative error at the $\mu + 2\sigma$ limit was found to be $12.6\%$ whereas the relative error with respect to the ensemble mean was found to be in the same range as the error from our deterministic model (see table \ref{tab:ensemble_error}).
\end{enumerate}
As a continuation to this work, we are exploring ways to extend the DeepONet model to generalize across different operating conditions. The operating ambient conditions, temperature and pressure can lead to different response from the engine. This could be accomplished by adding a new branch input and training with simulated data generated by varying the ambient temperature and pressure parameters in the Simulink model. 

\section*{Acknowledgement:} 
This research was conducted using computational resources and services at the Center for Computation and Visualization, Brown University.

\section*{Declarations}
\begin{itemize}
\item \textbf{Funding:} This study was funded by Cummins Inc.
\item \textbf{Competing interests:} Author Varun Kumar and Somdatta Goswami declare they have no financial interest. Author George Em Karniadakis has received research support from Cummins Inc.
\item \textbf{Ethical and informed consent for data used:} Not applicable.
\item \textbf{Data availability and access:} The datasets generated during and/or analysed during the current study are available from the corresponding author and with permission from Cummins Inc. on reasonable request post publication.
\item \textbf{Authors' contributions:} Varun Kumar was responsible for data generation, data processing, machine learning model design, coding, result interpretation, and material preparation. Somdatta Goswami was responsible for providing expertise in operator network design, material preparation and reviewing manuscript. Daniel Smith provided necessary guidance for data generation, problem setup and manuscript review. George Karniadakis provided critical feedback on manuscript and methods used in this study.
\end{itemize}


\bibliographystyle{elsarticle-num} 
\bibliography{references}

\begin{thebibliography}{10}
\expandafter\ifx\csname url\endcsname\relax
  \def\url#1{\texttt{#1}}\fi
\expandafter\ifx\csname urlprefix\endcsname\relax\def\urlprefix{URL }\fi
\expandafter\ifx\csname href\endcsname\relax
  \def\href#1#2{#2} \def\path#1{#1}\fi

\bibitem{AVLBoost}
AVL, {AVL Boost Engine simulation}, \url{https://www.avl.com/boost}, accessed:
  2022-08-08.

\bibitem{GTPower}
{Gamma Technologies}, {GT Power Engine Simulation},
  \url{https://www.gtisoft.com/gt-power/}, accessed: 2022-08-08.

\bibitem{RicardoWave}
{Ricardo Inc}, {WAVE 1D simulation},
  \url{https://software.ricardo.com/products/wave}, accessed: 2022-08-08.

\bibitem{hendricks1986compact}
E.~Hendricks, {A compact, comprehensive model of large turbocharged, two-stroke
  diesel engines}, {SAE Transactions} (1986) 820--834.

\bibitem{watson1984dynamic}
N.~Watson, {Dynamic turbocharged diesel engine simulator for electronic control
  system development}, 1984.

\bibitem{kimmich2005fault}
F.~Kimmich, A.~Schwarte, R.~Isermann, {Fault detection for modern Diesel
  engines using signal-and process model-based methods}, {Control Engineering
  Practice} 13~(2) (2005) 189--203.

\bibitem{wu2011mean}
H.~Wu, X.~Wang, R.~Winsor, K.~Baumgard, {Mean value engine modeling for a
  diesel engine with GT-Power 1D detail model}, Tech. rep., {SAE Technical
  Paper} (2011).

\bibitem{svard2010residual}
C.~Svard, M.~Nyberg, {Residual generators for fault diagnosis using computation
  sequences with mixed causality applied to automotive systems}, {IEEE
  Transactions on Systems, Man, and Cybernetics-Part A: Systems and Humans}
  40~(6) (2010) 1310--1328.

\bibitem{han1995turbulence}
Z.~Han, R.~D. Reitz, {Turbulence modeling of internal combustion engines using
  RNG $\kappa$-$\varepsilon$ models}, {Combustion Science and Technology}
  106~(4-6) (1995) 267--295.

\bibitem{goswami2020adaptive}
S.~Goswami, C.~Anitescu, T.~Rabczuk, {Adaptive fourth-order phase field
  analysis using deep energy minimization}, {Theoretical and Applied Fracture
  Mechanics} 107 (2020) 102527.

\bibitem{goswami2021physics}
S.~Goswami, M.~Yin, Y.~Yu, G.~E. Karniadakis, {A physics-informed variational
  DeepONet for predicting crack path in quasi-brittle materials}, {Computer
  Methods in Applied Mechanics and Engineering} 391 (2022) 114587.

\bibitem{lecun2015deep}
Y.~LeCun, Y.~Bengio, G.~Hinton, {Deep Learning}, {Nature} 521~(7553) (2015)
  436--444.

\bibitem{wahlstrom2011modelling}
J.~Wahlstr{\"o}m, L.~Eriksson, {Modelling diesel engines with a
  variable-geometry turbocharger and exhaust gas recirculation by optimization
  of model parameters for capturing non-linear system dynamics}, {Proceedings
  of the Institution of Mechanical Engineers, Part D: Journal of Automobile
  Engineering} 225~(7) (2011) 960--986.

\bibitem{Simulink}
J.~B. Dabney, T.~L. Harman, {Mastering Simulink}, Vol. 230, {Pearson/Prentice
  Hall Upper Saddle River}, 2004.

\bibitem{lu2021learning}
L.~Lu, P.~Jin, G.~Pang, Z.~Zhang, G.~E. Karniadakis, {Learning nonlinear
  operators via DeepONet based on the universal approximation theorem of
  operators}, Nature Machine Intelligence 3~(3) (2021) 218--229.

\bibitem{goswami2022physics}
S.~Goswami, A.~Bora, Y.~Yu, G.~E. Karniadakis, {Physics-Informed Neural
  Operators}, arXiv preprint arXiv:2207.05748 (2022).

\bibitem{lin2021operator}
C.~Lin, Z.~Li, L.~Lu, S.~Cai, M.~Maxey, G.~E. Karniadakis, Operator learning
  for predicting multiscale bubble growth dynamics, The Journal of Chemical
  Physics 154~(10) (2021) 104118.

\bibitem{cai2021deepm}
S.~Cai, Z.~Wang, L.~Lu, T.~A. Zaki, G.~E. Karniadakis, {DeepM\&Mnet: Inferring
  the electroconvection multiphysics fields based on operator approximation by
  neural networks}, Journal of Computational Physics 436 (2021) 110296.

\bibitem{chen1995universal}
T.~Chen, H.~Chen, {Universal approximation to nonlinear operators by neural
  networks with arbitrary activation functions and its application to dynamical
  systems}, {IEEE Transactions on Neural Networks} 6~(4) (1995) 911--917.

\bibitem{jin2022mionet}
P.~Jin, S.~Meng, L.~Lu, {MIONet: Learning multiple-input operators via tensor
  product}, arXiv preprint arXiv:2202.06137 (2022).

\bibitem{goswami2022neural}
S.~Goswami, D.~S. Li, B.~V. Rego, M.~Latorre, J.~D. Humphrey, G.~E.
  Karniadakis, {Neural operator learning of heterogeneous mechanobiological
  insults contributing to aortic aneurysms}, arXiv preprint arXiv:2205.03780
  (2022).

\bibitem{gal2016dropout}
Y.~Gal, Z.~Ghahramani, {Dropout as a Bayesian approximation: Representing model
  uncertainty in deep learning}, in: {International Conference on Machine
  Learning}, PMLR, 2016, pp. 1050--1059.

\bibitem{mcclenny2020self}
L.~McClenny, U.~Braga-Neto, Self-adaptive physics-informed neural networks
  using a soft attention mechanism, arXiv preprint arXiv:2009.04544 (2020).

\bibitem{kontolati2022influence}
K.~Kontolati, S.~Goswami, M.~D. Shields, G.~E. Karniadakis, {On the influence
  of over-parameterization in manifold based surrogates and deep neural
  operators}, arXiv preprint arXiv:2203.05071 (2022).

\bibitem{mackay1992practical}
D.~J. MacKay, {A practical Bayesian framework for backpropagation networks},
  {Neural Computation} 4~(3) (1992) 448--472.

\bibitem{jospin2022hands}
L.~V. Jospin, H.~Laga, F.~Boussaid, W.~Buntine, M.~Bennamoun, {Hands-on
  Bayesian neural networks—A tutorial for deep learning users}, {IEEE
  Computational Intelligence Magazine} 17~(2) (2022) 29--48.

\bibitem{psaros2022uncertainty}
A.~F. Psaros, X.~Meng, Z.~Zou, L.~Guo, G.~E. Karniadakis, {Uncertainty
  quantification in scientific machine learning: Methods, metrics, and
  comparisons}, arXiv preprint arXiv:2201.07766 (2022).

\bibitem{srivastava2014dropout}
N.~Srivastava, G.~Hinton, A.~Krizhevsky, I.~Sutskever, R.~Salakhutdinov,
  {Dropout: a simple way to prevent neural networks from overfitting}, {The
  Journal of Machine Learning Research} 15~(1) (2014) 1929--1958.

\bibitem{hastings1970monte}
W.~K. Hastings, {Monte Carlo sampling methods using Markov chains and their
  applications}, Oxford University Press, 1970.

\bibitem{bardenet2017markov}
R.~Bardenet, A.~Doucet, C.~C. Holmes, {On Markov chain Monte Carlo methods for
  tall data}, {Journal of Machine Learning Research} 18~(47) (2017).

\bibitem{neal2011mcmc}
R.~M. Neal, et~al., {MCMC using Hamiltonian dynamics}, {Handbook of Markov
  Chain Monte Carlo} 2~(11) (2011) 2.

\bibitem{blundell2015weight}
C.~Blundell, J.~Cornebise, K.~Kavukcuoglu, D.~Wierstra, {Weight uncertainty in
  neural network}, in: International Conference on Machine Learning, PMLR,
  2015, pp. 1613--1622.

\bibitem{hernandez2015probabilistic}
J.~M. Hern{\'a}ndez-Lobato, R.~Adams, {Probabilistic backpropagation for
  scalable learning of Bayesian neural networks}, in: {International Conference
  on Machine Learning}, PMLR, 2015, pp. 1861--1869.

\bibitem{blei2017variational}
D.~M. Blei, A.~Kucukelbir, J.~D. McAuliffe, {Variational inference: A review
  for statisticians}, {Journal of the American Statistical Association}
  112~(518) (2017) 859--877.

\bibitem{meng2022learning}
X.~Meng, L.~Yang, Z.~Mao, J.~del {\'A}guila~Ferrandis, G.~E. Karniadakis,
  {Learning functional priors and posteriors from data and physics}, {Journal
  of Computational Physics} 457 (2022) 111073.

\bibitem{raissi2019physics}
M.~Raissi, P.~Perdikaris, G.~E. Karniadakis, {Physics-informed neural networks:
  A deep learning framework for solving forward and inverse problems involving
  nonlinear partial differential equations}, {Journal of Computational Physics}
  378 (2019) 686--707.

\end{thebibliography}
\end{document}